\documentclass[conference]{IEEEtran}

\usepackage{amsfonts}
\usepackage{amssymb}
\usepackage{amsthm}
\usepackage{amsmath,amsfonts,amssymb}
\usepackage[dvips]{graphicx}
\usepackage{subfig}
\usepackage{verbatim}
\usepackage{setspace}
\usepackage{bm}
\usepackage{algorithmic} 
\usepackage[ruled,vlined]{algorithm2e}
\usepackage{cite}

\usepackage{changepage}
\usepackage{pdfpages}
\usepackage{color}

\usepackage{caption}

\IEEEoverridecommandlockouts

\begin{document}
\title{Movable Antennas for Wireless Communication: Opportunities and Challenges}
	
	\author{Lipeng Zhu, Wenyan Ma, and Rui Zhang
	\thanks{Lipeng Zhu (corresponding author) and Wenyan Ma are with National University of Singapore.}
	\thanks{Rui Zhang is with The Chinese University of Hong Kong, Shenzhen, Shenzhen Research Institute of Big Data, and National University of Singapore.}
}
\maketitle

\begin{abstract}
	Movable antenna (MA) technology is a recent development that fully exploits the wireless channel spatial variation in a confined region by enabling local movement of the antenna. Specifically, the positions of antennas at the transmitter and/or receiver can be dynamically changed to obtain better channel conditions for improving the communication performance. In this article, we first provide an overview of the promising applications for MA-aided wireless communication. Then, we present the hardware architecture and channel characterization for MA systems, based on which the variation of the channel gain with respect to the MA's position is illustrated. Furthermore, we analyze the performance advantages of MAs over conventional fixed-position antennas, in terms of signal power improvement, interference mitigation, flexible beamforming, and spatial multiplexing. Finally, we discuss the main design challenges and their potential solutions for MA-aided communication systems.
\end{abstract}

\section{Introduction}
Over the past few decades, the evolution from single-antenna or single-input single-output (SISO) to multi-antenna or multiple-input multiple-output (MIMO) has been a crucial trend in the development of wireless communication systems \cite{Wang2019massiveMIMO}. By leveraging the spatial multiplexing, interference suppression, beamforming, and diversity gains, MIMO technologies have significantly enhanced the capacity and reliability of wireless communications. However, existing communication systems cannot fully utilize the wireless channel spatial variation or so-called spatial degree of freedom (DoF) in the given regions where the transmitter (Tx) and receiver (Rx) reside due to the discrete form of antennas or antenna arrays deployed at fixed positions. To overcome this limitation, movable antenna (MA) technology was recently introduced to enable the local movement of antennas at the Tx/Rx in their specified regions \cite{zhu2022MAmodel}. With the aid of mechanical controllers and drivers, MAs can be continuously moved within a confined region, thereby fully exploiting the spatial DoF in the region for improving the wireless channel condition.

With the development of Internet of Things (IoT), future wireless networks will need to support the communications of massive IoT devices, for which a cost-effective solution is machine-type communication (MTC) \cite{Beyene2017narrowIoT,Dowhuszko2016delayMTC}. Typically, MTC is applicable for IoT devices that are deployed in confined areas at fixed locations, while the surrounding environment may change over time and results in slowly-varying wireless channels. In such low-mobility scenarios, narrow-band MTC usually has limited time and frequency diversity to utilize for improving the transmission reliability. As such, compared to conventional fixed-position antennas (FPAs), MA becomes a viable technology for achieving higher spatial diversity gains for slowly varying channels \cite{zhu2022MAmodel}. It is worth noting that spatial diversity can be easily obtained by the conventional antenna selection (AS) technique \cite{Sanayei2004antennas}. However, to achieve higher spatial diversity gains, the AS system needs to employ more antennas in an extended area from which a given number of antennas are selected based on their instantaneous channel state information (CSI). In contrast, the MA system can exploit the full spatial diversity in a given region with much fewer antennas or even a single antenna. Moreover, the antennas in AS systems can only be discretely deployed at fixed locations in a one-dimensional (1D) line or two-dimensional (2D) surface. In comparison, MAs can be flexibly moved in a three-dimensional (3D) region to fully exploit the channel variation therein. Thus, compared to AS, MA is a more cost-effective and efficient solution for exploring the spatial DoFs in confined regions.

\begin{figure*}[t]
	\begin{center}
		\includegraphics[width=14 cm]{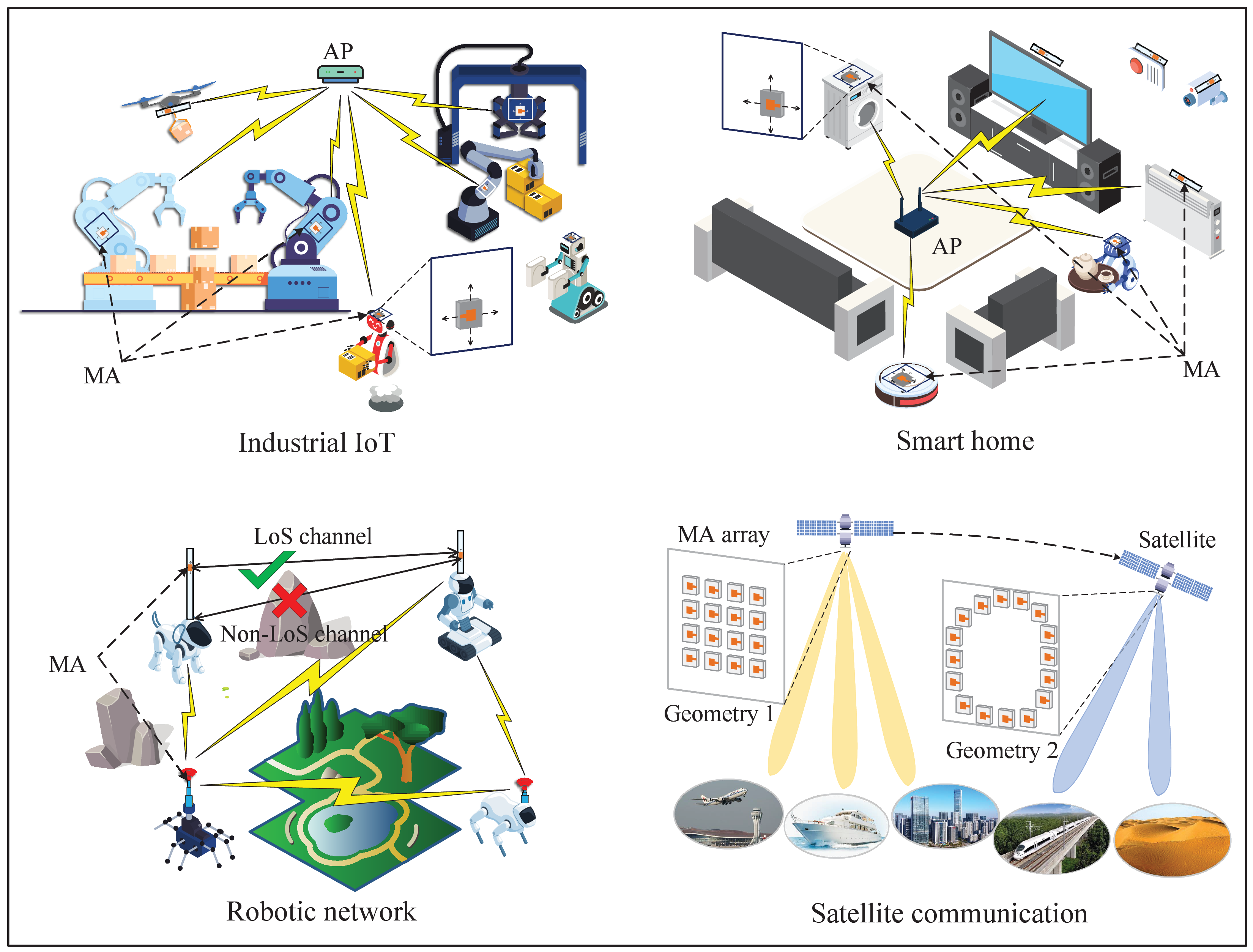}
		\caption{Typical applications for MA-aided wireless communication.}
		\label{fig:MA_Applications}
	\end{center}
\end{figure*}

Fig. \ref{fig:MA_Applications} illustrates the typical applications of MA-aided wireless communication, including 
\emph{(a) Industrial IoT}: The future industry will encompass automation and intelligence for reducing the need for human operations. In this context, a large number of machine-type terminals need to connect to their access points (APs) for reporting state information and receiving control instructions in real time. Since their surrounding propagation environment typically varies slowly and they are usually deployed at fixed locations or have low mobility, the MAs installed on them can help improve the wireless channel condition during a long channel coherence time, thereby enhancing their communication reliability with reduced latency;
\emph{(b) Smart home}: Future smart homes are expected to support wireless connectivity and information sharing between different types of devices, such as household appliances, domestic robots, and indoor sensors, to realize various functions. As the number of wireless devices increases, the bandwidth allocated to each device is limited, and the locations of such devices are usually fixed or change slowly over time. As a result, if the channel conditions between the AP and such devices deteriorate, conventional time/frequency diversity techniques are not effective. In such scenarios, mounting MAs on these devices can help increase the channel power without the need of installing additional antennas; 
\emph{(c) Robotic network}: In such networks, multiple robots may transmit/receive signals simultaneously in a given bandwidth, thus causing undesired interference which can degrade their communication performance. The integration of MAs to robot transceivers can significantly improve their spatial interference suppression capability, even with small number of antennas. For instance, the MAs can be moved to positions which achieve high channel gains between a pair of desired transceivers while minimizing the interference from/to other undesired Tx/Rx. Moreover, if the wireless link between two nodes is blocked by obstacles, the altitude of MAs can be adjusted to establish a line-of-sight (LoS) channel; and 
\emph{(d) Satellite communication}: In practice, satellites are usually equipped with large-scale FPAs for synthesizing narrow beams to compensate for the high path loss due to the long signal-propagation distance with ground users. Particularly, low-earth orbit (LEO) satellites may travel and serve different ground areas which may have distinct user distributions and beam coverage requirements \cite{Kodheli2020satell}. However, FPA arrays have a fixed geometry once manufactured, which can only adapt to the varying channel conditions and communication requirements via analog/digital beamforming. In contrast, the geometry of an MA array can be reconfigured and thus more flexible beam patterns can be obtained by jointly optimizing the positions and beamforming weights of MAs to enhance the beam coverage for ground users. Although the satellite usually maintains a high moving speed, the steering angles for covering terrestrial regions vary slowly over time due to its high altitude relative to the ground. Moreover, the orbit of the satellite is fixed and follows a periodic pattern, thus rendering the optimization and implementation of MAs' positions for satellite practically feasible.

Considering the fertile opportunities and promising applications of MAs for future wireless communication systems, this article aims to provide an overview of the fundamentals for MA-aided communications as well as the main challenges in their implementation. To this end, we first introduce the architecture of MA systems and their channel characterization. We then analyze the major performance gains provided by antenna movement in terms of signal power improvement, interference mitigation, flexible beamforming, and spatial multiplexing. Finally, we discuss the technical challenges in designing and implementing MA-aided communication systems and point out potential solutions to tackle them.

\section{Architecture and Channel Characterization}
\begin{figure}[t]
	\begin{center}
		\includegraphics[width=8.8 cm]{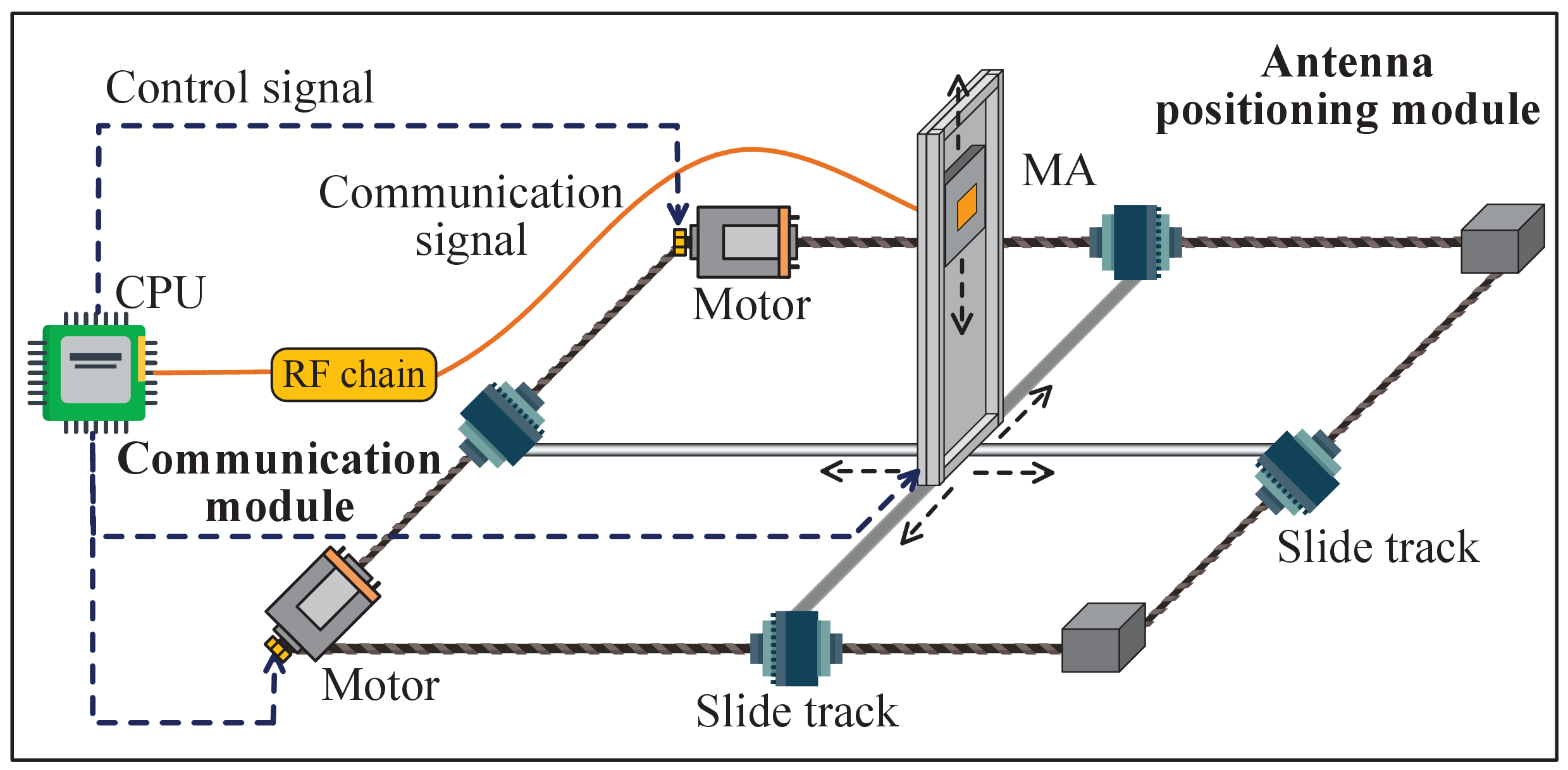}
		\caption{An architecture for the MA-mounted Tx/Rx.}
		\label{fig:MA_Architecture}
	\end{center}
\end{figure}

Fig. \ref{fig:MA_Architecture} illustrates an example architecture for MA-mounted Tx/Rx, which comprises of a communication module and an antenna positioning module. Specifically, the communication module is similar to that of conventional FPA systems, whereas the MA is connected to the radio frequency (RF) chain via a flexible cable to support antenna movement. The central processing unit (CPU) is used for digital signal processing as well as controlling the antenna position. For the antenna positioning module, the MA is installed on a 3D mechanical slide which is driven by step motors \cite{Zhuravlev2015experi,Li2022movingAnte}. After receiving the control signal from the CPU, the motors can cooperatively perform the corresponding steps to relocate the antenna to a target position, with desired accuracy (e.g., in tenth of the wavelength). Besides, the orientation change of the MA at a given position aided by a servo motor offers three additional DoFs for antenna movement. As such, six-dimensional (6D) DoFs can be generally explored for MA systems in total. For multi-MA systems, each MA can be driven by an independent positioning module and moved in a separate sub-region. In addition to the motor slide-based architecture, there are other ways to implement MAs by, e.g., enabling the rotation of small-scale antennas along a circular track \cite{Basbug2017design} or moving large-scale antennas with the aid of vehicles \cite{Haupt2013reconfig}. Besides, the microelectromechanical systems (MEMS)-integrated antenna \cite{balanis2008mems} and the liquid antenna (or termed as fluid antenna) \cite{paracha2019liquid} are two alternative solutions for implementing MAs with high positioning accuracy in portable devices with compact components. Preliminary studies have shown that MA systems can achieve a considerable performance gain over FPA systems by moving antennas in regions with size of only a few wavelengths \cite{zhu2022MAmodel,ma2022MAmimo,zhu2023MAmultiuser}. With the trend of wireless communications migrating to higher frequency, such as millimeter-wave and terahertz frequency bands \cite{xiao2022mmWaveUAV}, the decreasing wavelength makes it more feasible to implement MAs in a small area. 

\begin{figure*}
	\begin{center}
	\includegraphics[width=18 cm]{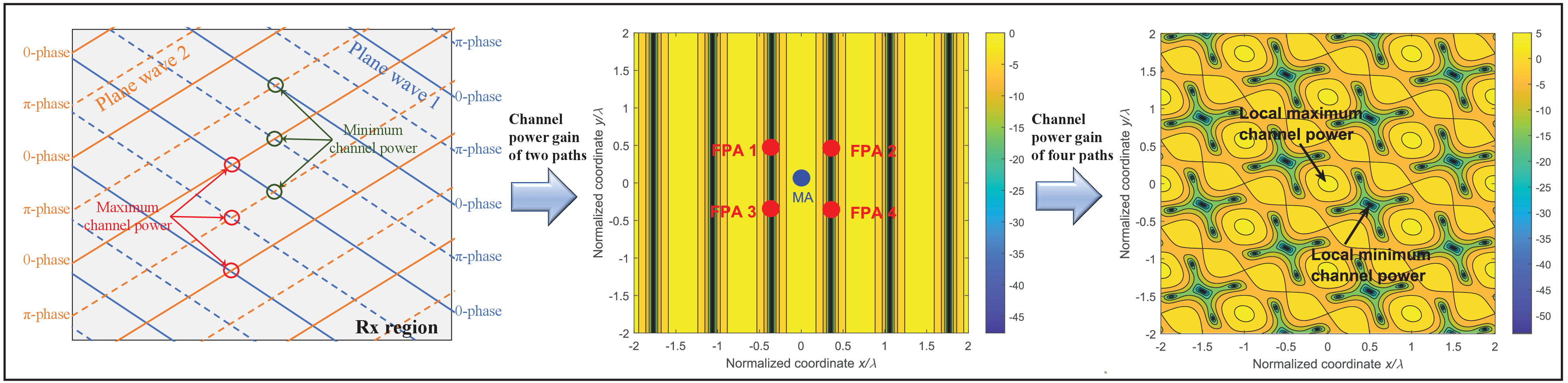}\\
	\caption{Illustration of the variation of channel power gain (in dB) in the Rx region.}
	\label{fig:channel}
	\end{center}
\end{figure*}
	
For MA systems, a fundamental problem is to develop a general model that characterizes the channel response between Tx-side and Rx-side MAs located at different positions in their respective regions. Since the distance between the Tx and Rx is practically much longer than the size of MA's moving region, the far-field channel condition holds at both the Tx and Rx sides. Under this condition, the uniform plane wave model can be used to characterize the response of each channel path between the Tx and Rx. In particular, the angle of departure (AoD), the angle of arrival (AoA), and the amplitude of the complex coefficient associated with each channel path between the Tx and Rx can be assumed to be constant over their residing regions, regardless of the positions of the MAs therein \cite{zhu2022MAmodel}. However, the phase of each path coefficient can vary rapidly with antenna position due to the change in signal propagation distance with respect to the signal wavelength. From the above, it can be inferred that the multi-path channel characterization between the residing regions of Tx and Rx only relies on the parameters of different channel paths, which do not scale with the region sizes. This fact will be exploited later for efficient CSI acquisition for MAs. 

Next, we illustrate the channel response in the Rx region as an example, where the position of Tx's antenna is assumed to be fixed. As illustrated in Fig. \ref{fig:channel}, two plane waves arriving from different channel paths with distinct AoAs combine in the Rx region. The two waves are constructively superimposed at positions (marked by red circles) where they share the same phase, leading to the maximum channel power gain. Conversely, in positions where the two waves have opposite phases (marked by green circles), the coefficients of the two channel paths cancel each other, resulting in the minimum channel power gain. Due to the distinct wavefront directions, the phases of the two plane waves have different periodic variation patterns over the region, which cause the channel gain to vary with the antenna position. Fig. \ref{fig:channel} also shows the channel power gain of two unit-power channel paths in a square Rx region. As can be observed, the channel power gain exhibits a periodic variation, which depends on the AoA difference of the two channel paths. Moreover, the gap between the maximum and minimum channel power gains can be substantial (over 40 dB in this example) in a small region with size of only several wavelengths. As the number of channel paths increases, the channel gain variation becomes more pronounced in the spatial region, while its periodic behavior becomes less explicit because of the intricate superposition of more channel paths' coefficients. For the case of 4 unit-power channel paths with different AoAs shown in Fig. \ref{fig:channel}, it can be observed that the channel power gain exhibits considerable variation in the spatial region, even if the antenna moves within a sub-wavelength scale. 

\section{Performance Advantages of MA over FPA}
In this section, we analyze the performance advantages of MA over FPA from four main perspectives, which are signal power improvement, interference mitigation, flexible beamforming, and spatial multiplexing. For simplicity, we focus on the MAs at Rx to illustrate their advantages, while similar results can be inferred for Tx-side MAs as well as MAs at both Tx and Rx.

\begin{figure*}
	\begin{center}
	\includegraphics[width=16 cm]{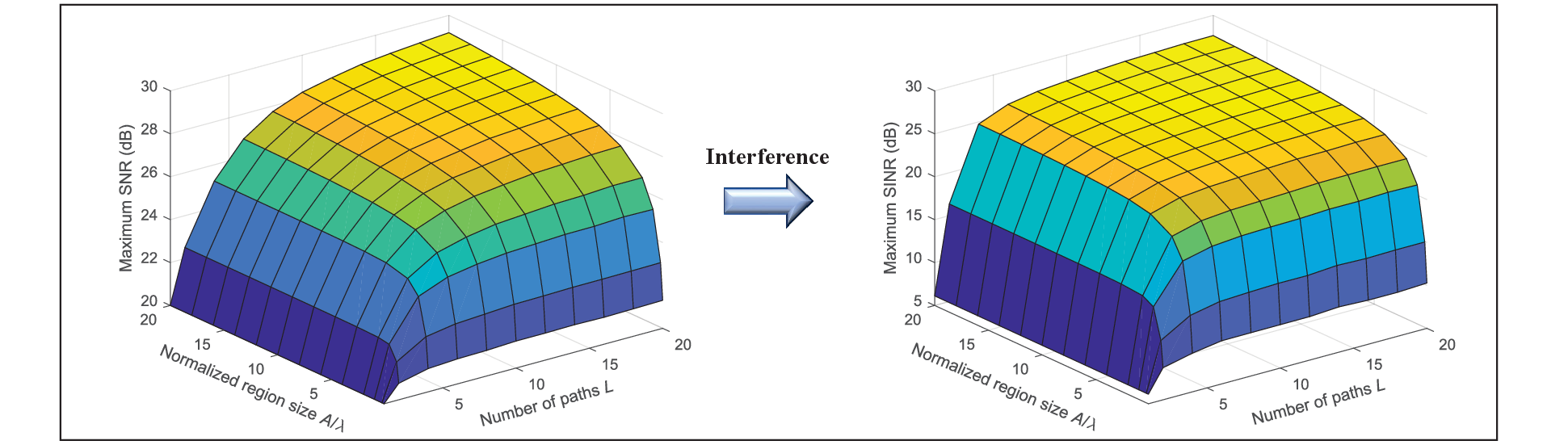}
	\caption{The expected values of the maximum SNR and SINR achieved by a single Rx-side MA versus the number of channel paths and the size of Rx region.}
	\label{fig:SNR&SINR}
	\end{center}
\end{figure*}
	
\subsection{Signal Power Improvement}
The position optimization of MAs can significantly increase the received signal power by improving the channel gains, especially for a large number of channel paths and a large size of Rx region \cite{zhu2022MAmodel}.  As shown in Fig. \ref{fig:channel}, for a given channel gain map, if multiple FPAs are all located at positions with deep fading channels, the communication performance cannot be guaranteed even AS or maximal ratio combining (MRC) is employed at the Rx. In contrast, the MA can be easily moved to a position with higher channel gain such that the received signal power can be increased. 

To evaluate the performance gain by improving the received signal power, Fig. \ref{fig:SNR&SINR} shows the maximum signal-to-noise ratio (SNR) achieved by a single Rx-side MA versus the number of channel paths ($L$) and the size of Rx region under stochastic channels. The setup assumes a single FPA at the Tx and a single MA at the Rx moving in a square region of size $A \times A$, with $\lambda$ denoting the wavelength. The coefficient of each receive channel path is modeled as an independent and identically distributed (i.i.d.) circularly symmetric complex Gaussian (CSCG) random variable with mean zero and variance $1/L$ \cite{zhu2022MAmodel}. The azimuth and elevation AoAs of each receive channel path are assumed to follow a uniform distribution within the range of $[-\pi/2, \pi/2]$. For each channel realization, due to the variation of the channel gain over the Rx region, the receive SNR changes with the MA position, where the best position is selected for achieving the maximum SNR over the whole Rx region. Each point in Fig. \ref{fig:SNR&SINR} is the average result over $10^4$ random channel realizations. The average receive SNR at the (fixed) reference point in the square region over is set as 20 dB. The results show that the expected maximum SNR of the Rx-side MA increases with both the number of channel paths and the size of Rx region, e.g., about 10 dB increase in SNR for $A=20 \lambda$ and $L=20$. If the channel power is dominant by an LoS path (which is a special case of non-i.i.d. channel coefficients), the performance gain provided by antenna movement may decrease because the small-scale fading of the wireless channel becomes less pronounced over the spatial region. It is worth noting that the above comparison is based on stochastic channels (under random AoAs and path coefficients) and considers the average SNR; while as shown in Fig. \ref{fig:channel}, the worst-case SNR with FPA, which corresponds to the position with local minimum channel power, can be much lower than that with a single MA.

\subsection{Interference Mitigation}
In addition to increasing the power of desired signals, the position optimization of MAs can also help reduce interference power. For instance, as shown in Fig. \ref{fig:channel}, the MA can be placed at the position with the local minimum channel power gain from the interfering Tx, which requires only a sub-wavelength movement of the antenna but leads to a significant decrease (tens of dB) in the interference power. As such, the interference can be efficiently mitigated even with a single antenna at the Rx.

To evaluate the performance gain by mitigating interference, Fig. \ref{fig:SNR&SINR} also illustrates the expected values of the maximum SINR achieved by a single Rx-side MA versus the number of channel paths (from the desired Tx) and the size of Rx region. In addition to the desired Tx, the impact of an interfering Tx with an average interference-to-noise ratio of 20 dB to the reference point in the Rx region is considered, where the stochastic channel setup is similar to that of the desired Tx. For each of the $10^4$ random channel realizations, the receive SINR varies with the position of the MA, which is chosen to be the best location for achieving the maximum SINR over the whole Rx region. As the number of channel paths and the size of Rx region increase, a more favorable position can be found in the Rx region for minimizing the interference power as well as maximizing the desired signal power. Interestingly, for a sufficiently large number of channel paths and large size of the Rx region, the expected maximum SINR is close to the expected maximum SNR. This implies that for the SINR-maximum position of the Rx-side MA, the interference is significantly suppressed with almost no loss of the desired signal power. It is worth noting that the complete interference suppression is achieved by optimally positioning a single MA only, without the need of multi-antenna interference cancellation/nulling.

\begin{figure*}
	\begin{center}
		\includegraphics[width=16 cm]{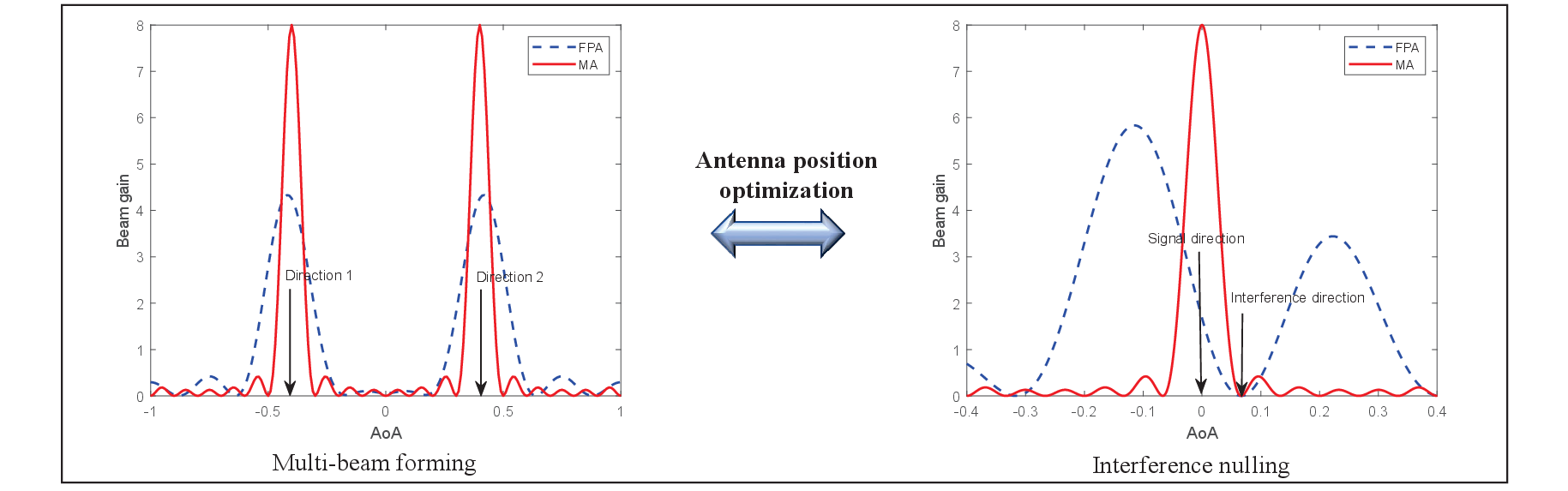}
		\caption{The comparison of beam patterns between FPA and MA arrays.}
		\label{fig:beamforming}
	\end{center}
\end{figure*}

\subsection{Flexible Beamforming}
Conventional FPA arrays usually have a fixed geometry once manufactured, which limits their beamforming performance. In contrast, the geometry of an MA array can be reconfigured, such that more flexible beamforming can be achieved by jointly designing the positions and beamforming weights of multiple antennas. For example, when implementing multi-beam forming, conventional FPA arrays with fixed antenna spacing experience a loss of array gains. This is because the steering vectors, which depend on the array geometry, are approximately orthogonal in the beamspace if the AoAs are far apart in the angle domain; while a beamforming vector cannot achieve a high correlation with two (approximately) orthogonal steering vectors simultaneously. In contrast, the geometry of an MA array can be reshaped to increase the correlation between the steering vectors of multiple target AoAs, thus enabling multi-beam forming with much less loss of the individual array gains in different directions. In Fig. \ref{fig:beamforming}, we compare the beam patterns of linear FPA and MA arrays for multi-beam forming towards two directions with AoAs (in the cosine domain) equal to $-0.4$ and $0.4$, respectively. We set the number of antennas to 8 for both arrays, with the FPA employing a uniform linear array (ULA) of $\lambda/2$ spacing. In particular, the FPA-ULA employs the near-optimal multi-beam forming solution proposed in \cite{xiao2022mmWaveUAV}, while the MA array increases the spacing between adjacent antennas to $1.25\lambda$ and uses its steering vector corresponding to AoA of $0.4$ for beamforming. As can be observed, the ULA with fixed antennas can only achieve approximately half of the full array gain for each of the two desired directions. However, by adjusting the positions of the antenna elements, the MA array can achieve full array gains for both desired directions.

Another benefit of MA arrays is their ability to synthesize beam patterns with more flexible null directions. For FPA arrays, when using a steering vector as the beamforming vector to maximize the array gain over an AoA, the interference from directions close to this desired AoA cannot be efficiently suppressed. Conversely, if the beam gain of the interference direction within the beam width is nulled or minimized, the array gain over the desired AoA may be significantly reduced. This issue stems from the high correlation between the steering vectors of AoAs within the beam width for a given FPA array geometry. MA arrays can overcome this limitation by allowing flexible reconfiguration of the array geometry, thus reducing the correlation between the steering vectors of the desired and interference AoAs. As a result, null steering of the beam over interference direction may not dramatically sacrifice the array gain of desired AoA. In Fig. \ref{fig:beamforming}, the signal and interference AoAs in the cosine domain are set to $0$ and $1/15$, respectively. For the ULA with 8 FPA elements, the interference nulling by zero forcing results in a significant loss of the array gain. In contrast, by optimizing the positions of the MAs (where the spacing between adjacent antennas is adjusted to $15\lambda/8$), the MA array can achieve full array gain over the signal direction while completely nulling the signal from the interference direction. Note that the beamforming gains in Fig. \ref{fig:beamforming} are attained by only adjusting the spacing between adjacent MAs. More flexible beamforming can be achieved by jointly optimizing the positions and beamforming weights of all antenna elements, which will be an interesting topic for future research.

\subsection{Spatial Multiplexing}
The capacity of a MIMO system, which depends on the singular values of the channel matrix, can be improved by reshaping the channel matrix through MA position optimization \cite{ma2022MAmimo}. In the low-SNR regime, the single-stream beamforming strategy is optimal, where all transmit power is allocated to the strongest eigenchannel of the MIMO channel. In this case, MA positioning can be optimized to increase the maximum singular value of the channel matrix. In contrast, in the high-SNR regime, the transmit power is allocated to multiple eigenchannels based on the water-filling principle, and thus MAs' positions should be optimized to balance the singular values of the channel matrix in order to maximize the capacity, rather than simply maximizing the channel power gain of each antenna.

\begin{figure}[t]
	\begin{center}
		\includegraphics[width=8.8 cm]{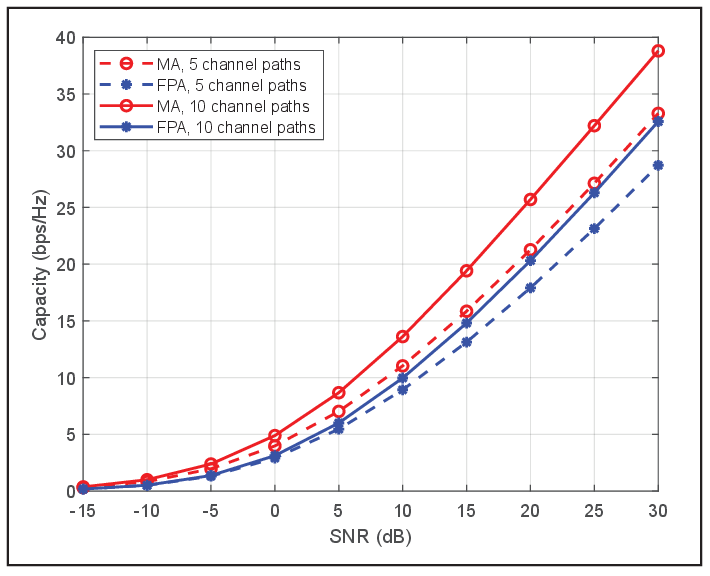}
		\caption{MIMO capacity of MA and FPA systems versus the average SNR, with the size of Rx region being $3\lambda \times 3\lambda$.}
		\label{fig:MIMO}
	\end{center}
\end{figure}

To evaluate the MIMO capacity gain provided by MAs, we plot the capacity of MA-MIMO and FPA-MIMO systems versus the average SNR at the Rx, which is defined as the ratio of the average received signal power to noise power between a single Tx-side FPA with transmit power $P_{\mathrm{t}}$ and a single Rx-side FPA with noise power $\sigma^{2}$. For FPA-MIMO systems, the FPAs are employed at both Tx and Rx. In contrast, for MA-MIMO systems, the FPAs are utilized at Tx with $\lambda/2$ spacing and the MAs are used at Rx. We consider systems with $N=4$ and $M=4$ transmit/receive antennas, respectively. To avoid the coupling effect, the positions of two MAs should be separated by no less than $\lambda/2$. In the simulation, the CSI is assumed to be known at the Rx only, and thus an identity matrix scaled by $P_{\mathrm{t}}/N$ is employed as the transmit covariance matrix. The position of each Rx-side MA is sequentially searched to maximize the channel capacity between Tx and Rx, with the positions of the other Rx-side MAs being fixed. As shown in Fig. \ref{fig:MIMO}, the MA-MIMO systems achieve higher capacity than their corresponding FPA-MIMO systems. Moreover, as the number of channel paths increases, the small-scale fading becomes more prominent in the Rx region, and thus the capacity gain of MA-MIMO over FPA-MIMO also increases.

\section{Challenges and Solutions}
\subsection{Channel Estimation}
To fully leverage the performance gains in the spatial domain, it is crucial for MA-aided systems to obtain accurate CSI at Tx/Rx. In general, a complex-valued channel map between Tx and Rx regions needs to be constructed, which includes the channel response from each point in the Tx region to each point in the Rx region. If the size of the Tx/Rx region for antenna moving is small, exhaustive antenna movement can be used for measuring the channel map over the entire Tx/Rx region. However, for larger Tx/Rx regions, exhaustive-movement-based channel mapping may result in prohibitively high training overhead because the Tx/Rx MAs have to visit all possible locations in their corresponding regions to perform channel measurement. This also incurs high energy consumption for mechanically moving the antennas.

To reduce the overhead for channel measurement, it is more efficient to estimate the field-response information (FRI) in the angle domain, including the AoDs, the AoAs, and the coefficients of multiple channel paths at the Tx/Rx side, for reconstructing the channel map between them \cite{zhu2022MAmodel}. In this way, the time overhead for channel estimation only depends on the number of channel paths, instead of the size of Tx/Rx region in the case of exhaustive search. By moving MAs to a sufficient number of locations for channel measurement, the FRI of multipath components can be resolved by using the Fourier transform relationship between the channel response in the location domain and the path response in the angle domain \cite{zhu2022MAmodel}. To this end, sparse signal recovery algorithms, such as compressed sensing (CS), can be employed, where the measurement locations of MAs determine the measurement matrix of the CS algorithm and thus significantly influence the estimation accuracy. Additionally, if Tx and Rx are fixed at confined regions or have low mobility, the AoDs/AoAs usually exhibit slow time-varying characteristics compared to the complex coefficients of the channel paths. This thus motivates the employment of a two time-scale FRI estimation strategy. Specifically, the AoDs/AoAs can be estimated in a large time-scale, while the path coefficients are estimated in a small time-scale with the AoD/AoA information as a priori knowledge, such that the channel estimation overhead can be further reduced. Developing more efficient channel estimation and channel map reconstruction algorithms for MA-aided systems is a timely and important research problem to solve in future work. 
	
	
\subsection{MA Position Optimization}
The performance gain of MA systems over conventional FPA systems is attributed to their flexible antenna positioning, which has been substantiated by our analysis and simulation results presented previously. However, finding the optimal antenna positions that yield the best communication performance poses a critical challenge because the channel response is a highly non-linear function of the MAs' positions. For the case with perfect CSI, local optimization methods, such as gradient descent/ascent and successive convex approximation (SCA), can be used to pursue suboptimal solutions for MAs' positions in the continuous Tx/Rx region. If the number of MAs is large, alternating optimization (AO) can be leveraged to reduce the computational complexity. For instance, the position of each Tx-/Rx-side MA was alternatively optimized in \cite{ma2022MAmimo} by using the SCA technique. Besides, the gradient descent method was utilized in \cite{zhu2023MAmultiuser} to jointly optimize the positions of MAs for all users.

For the case without perfect CSI, a straightforward approach would be to perform an exhaustive search over the candidate MA positions in the Tx/Rx region and select the ones achieving the best communication performance. However, to guarantee optimality, the number of candidate positions for MAs should be sufficiently large, which incurs prohibitively high time and energy overhead, especially for a large number of MAs and a large size of regions for antenna moving. Considering the above challenge, machine learning methods, e.g., deep neural network and reinforcement learning, may be efficient approaches for MA position optimization. Specifically, the positions of MAs can be optimized without knowing the instantaneous channel map, but instead by leveraging decisions provided by machine learning algorithms. For example, the neural network may learn a very large number of channel maps between the transceivers and extract the environment information in an offline manner. Then, by sampling a finite number of positions in the Tx/Rx region and using their channel measurements as the input, the well-trained neural network can output the favorable positions for MAs to improve the communication performance, without any explicit FRI estimation and channel map reconstruction. Thus, machine learning approaches for designing MA-aided communication systems will be an interesting topic for future research.

\section{Conclusions}
In this article, we provided an overview of the applications, fundamentals, challenges, as well as solutions for MA-aided communication systems. Compared to conventional FPAs, MA-aided communication can fully exploit the wireless channel spatial variation in confined regions, leading to improved signal power, suppressed interference, flexible beamforming, and enhanced spatial multiplexing performance in a cost-effective manner. However, MA systems also face practical challenges, such as channel estimation and antenna position optimization. More research efforts are thus needed to tackle these challenging problems and make MA-aided communication a practically useful technology for future wireless networks.
	
\section*{Acknowledgements}
This work is supported in part by Ministry of Education, Singapore under Award T2EP50120-0024, Advanced Research and Technology Innovation Centre (ARTIC) of National University of Singapore under Research Grant R-261-518-005-720, and The Guangdong Provincial Key Laboratory of Big Data Computing.

\bibliographystyle{IEEEtran} 
\bibliography{IEEEabrv, ref_zhu}

\begin{thebibliography}{10}
\providecommand{\url}[1]{#1}
\csname url@samestyle\endcsname
\providecommand{\newblock}{\relax}
\providecommand{\bibinfo}[2]{#2}
\providecommand{\BIBentrySTDinterwordspacing}{\spaceskip=0pt\relax}
\providecommand{\BIBentryALTinterwordstretchfactor}{4}
\providecommand{\BIBentryALTinterwordspacing}{\spaceskip=\fontdimen2\font plus
\BIBentryALTinterwordstretchfactor\fontdimen3\font minus
  \fontdimen4\font\relax}
\providecommand{\BIBforeignlanguage}[2]{{%
\expandafter\ifx\csname l@#1\endcsname\relax
\typeout{** WARNING: IEEEtran.bst: No hyphenation pattern has been}%
\typeout{** loaded for the language `#1'. Using the pattern for}%
\typeout{** the default language instead.}%
\else
\language=\csname l@#1\endcsname
\fi
#2}}
\providecommand{\BIBdecl}{\relax}
\BIBdecl

\bibitem{Wang2019massiveMIMO}
M.~Wang, F.~Gao, S.~Jin, and H.~Lin, ``An overview of enhanced massive {MIMO}
  with array signal processing techniques,'' \emph{IEEE J. Sel. Top. Sign.
  Proces.}, vol.~13, no.~5, pp. 886--901, Sep. 2019.

\bibitem{zhu2022MAmodel}
L.~Zhu, W.~Ma, and R.~Zhang, ``Modeling and performance analysis for movable
  antenna enabled wireless communications,'' \emph{arXiv preprint: 2210.05325},
  \url{https://arxiv.org/abs/2210.05325}, accessed on 11 Oct. 2022.

\bibitem{Beyene2017narrowIoT}
Y.~D. Beyene, R.~Jantti, K.~Ruttik, and S.~Iraji, ``On the performance of
  narrow-band internet of things ({NB-IoT}),'' in \emph{Proc. IEEE Wireless
  Commun. Networking Conf.}, Mar. 2017, pp. 1--6.

\bibitem{Dowhuszko2016delayMTC}
A.~A. Dowhuszko \emph{et~al.}, ``Performance of quantized random beamforming in
  delay-tolerant machine-type communication,'' \emph{IEEE Trans. Wireless
  Commun.}, vol.~15, no.~8, pp. 5664--5680, Aug. 2016.

\bibitem{Sanayei2004antennas}
S.~Sanayei and A.~Nosratinia, ``Antenna selection in {MIMO} systems,''
  \emph{IEEE Commun. Mag.}, vol.~42, no.~10, pp. 68--73, Oct. 2004.

\bibitem{Kodheli2020satell}
O.~Kodheli \emph{et~al.}, ``Satellite communications in the new space era: A
  survey and future challenges,'' \emph{IEEE Commun. Surveys Tuts.}, vol.~23,
  no.~1, pp. 70--109, 1st Quart. 2021.

\bibitem{Zhuravlev2015experi}
A.~Zhuravlev \emph{et~al.}, ``Experimental simulation of multi-static radar
  with a pair of separated movable antennas,'' in \emph{Proc. IEEE
  International Conf. Microwaves Commun. Antennas Electron. Syst.}, Nov. 2015,
  pp. 1--5.

\bibitem{Li2022movingAnte}
X.~Li \emph{et~al.}, ``Using a moving antenna to improve {GNSS/INS} integration
  performance under low-dynamic scenarios,'' \emph{IEEE Trans. Intell.
  Transport. Syst.}, vol.~23, no.~10, pp. 17\,717--17\,728, Oct. 2022.

\bibitem{Basbug2017design}
S.~Basbug, ``Design and synthesis of antenna array with movable elements along
  semicircular paths,'' \emph{IEEE Antennas Wireless Propagat. Lett.}, vol.~16,
  pp. 3059--3062, Oct. 2017.

\bibitem{Haupt2013reconfig}
R.~L. Haupt and M.~Lanagan, ``Reconfigurable antennas,'' \emph{IEEE Antennas
  Propagat. Mag.}, vol.~55, no.~1, pp. 49--61, Feb. 2013.

\bibitem{balanis2008mems}
B.~Pan, J.~Papapolymerou, and M.~M. Tentzeris, ``{MEMS} integrated and
  micromachined antenna elements, arrays, and feeding networks,'' \emph{Modern
  Antenna Handbook, John Wiley \& Sons}, pp. 829--865, Sep. 2008.

\bibitem{paracha2019liquid}
K.~N. Paracha \emph{et~al.}, ``Liquid metal antennas: Materials, fabrication
  and applications,'' \emph{Sensors}, vol.~20, no.~1, pp. 1--26, Dec. 2019.

\bibitem{ma2022MAmimo}
W.~Ma, L.~Zhu, and R.~Zhang, ``{MIMO} capacity characterization for movable
  antenna systems,'' \emph{IEEE Trans. Wireless Commun.},
  \url{https://ieeexplore.ieee.org/abstract/document/10243545}, accessed on 7
  Sep. 2023, DOI: 10.1109/TWC.2023.3307696.

\bibitem{zhu2023MAmultiuser}
L.~Zhu, W.~Ma, B.~Ning, and R.~Zhang, ``Movable-antenna enhanced multiuser
  communication via antenna position optimization,'' \emph{arXiv preprint:
  2302.06978}, \url{https://arxiv.org/abs/2302.06978}, accessed on 14 Feb.
  2023.

\bibitem{xiao2022mmWaveUAV}
Z.~Xiao \emph{et~al.}, ``A survey on millimeter-wave beamforming enabled {UAV}
  communications and networking,'' \emph{IEEE Commun. Surveys Tuts.}, vol.~24,
  no.~1, pp. 557--610, 1st Quart. 2022.

\end{thebibliography}

{Lipeng Zhu} (zhulp@nus.edu.sg) is a Research Fellow with the Department of Electrical and Computer Engineering, National University of Singapore.
	
{Wenyan Ma} (wenyan@u.nus.edu) is a Ph.D. candidate with the Department of Electrical and Computer Engineering, National University of Singapore.

{Rui Zhang} [F'17] (elezhang@nus.edu.sg) is a professor with the School of Science and Engineering, The Chinese University of Hong Kong, Shenzhen, and Shenzhen Research Institute of Big Data. He is also with the ECE Department of National University of Singapore.

\end{document}